\documentclass{ws-procs9x6}
\usepackage{graphicx}
\usepackage{amsmath,amssymb,amsfonts}
\usepackage{latexsym}

\begin{document}

\title{Braneworld cosmology: sneutrino inflation and
leptogenesis \footnote{ \uppercase{T}alk given by
\uppercase{N.M.C.S.} at the \uppercase{XV ENAA}, \uppercase{L}isbon,
\uppercase{P}ortugal.}}

\author{N.~M.~C.~Santos\footnote{\uppercase{P}resently at \uppercase{I}nstitut
f\"{u}r \uppercase{T}heoretische \uppercase{P}hysik,
\uppercase{U}niversit\"{a}t \uppercase{H}eidelberg,
\uppercase{P}hilosophenweg 16, 69120 \uppercase{H}eidelberg,
\uppercase{G}ermany. \uppercase{N.M.C.S.} acknowledges the support
of \uppercase{F}unda\c c\~ao para a \uppercase{C}i\^encia e a
\uppercase{T}ecnologia (\uppercase{FCT}, \uppercase{P}ortugal)
under the grant \uppercase{SFRH/BD}/4797/2001.}, ~M.~C.~Bento and
~R.~Gonz\'{a}lez~Felipe}

\address{Departamento de F\'{\i}sica and Centro de F\'{\i}sica
Te\'orica de Part\'{\i}culas, \\
Instituto Superior T\'{e}cnico, Av. Rovisco Pais, \\
1049-001 Lisboa, Portugal\\
E-mail: n.santos@thphys.uni-heidelberg.de}

\maketitle

\abstracts{ Modifications to the Friedmann equation in brane
cosmology can have important implications for early universe
phenomena such as inflation and the generation of the baryon
asymmetry. In the framework of braneworld cosmology, we discuss a
mechanism for baryogenesis via leptogenesis in the supersymmetric
context, where the sneutrino is responsible for both inflation and
the generation of the baryon asymmetry in the universe.}

Today there is a wide consensus that the early universe underwent
a period of cosmological inflation, responsible not only for the
observed flatness, homogeneity and isotropy of the present
universe, but also for the origin of the density fluctuations. At
the end of inflation, the universe must have been subsequently
reheated to become a high-entropy and radiation-dominated
universe. Such a reheating process could occur, for instance,
through the coherent oscillations of the inflaton field about the
minimum of the potential. During this process, the inflaton decays
into ordinary particles, which then scatter and thermalise. The
right abundance of baryons must have also been created after
inflation. In fact, the most recent Wilkinson Microwave Anisotropy
Probe (WMAP) results and big bang nucleosynthesis (BBN) analysis
of primordial deuterium abundance imply\cite{WMAP} ${\mbox\,
\eta_B \equiv (n_B-n_{\bar{B}})/n_\gamma=(6.1 \pm 0.3)\times
10^{-10}}$, for the baryon-to-photon ratio of number densities.

In this talk, we present a minimal supersymmetric seesaw scenario
where the lightest singlet sneutrino field not only plays the role
of the inflaton but also produces a lepton asymmetry through its
direct decays, during the nonconventional era in the braneworld
scenario\cite{Bento:2004pz}. There are other baryogenesis
scenarios in which the braneworld modifications can have important
implications, e.g., the recently proposed gravitational
baryogenesis mechanism\cite{Davoudiasl:2004gf}.

Whilst theories formulated in extra dimensions have been around
since the early twentieth century, recent developments in string
theory have opened up the possibility that our universe could be a
1+3-surface - the brane - embedded in a higher-dimensional
space-time, called the bulk, with standard model particles and
fields trapped on the brane while gravity is free to access the
bulk\cite{Maartens:2003tw}. A remarkable feature of brane
cosmology (BC) is the modification of the expansion rate of the
universe before the BBN era. In the so-called Randall-Sundrum II
braneworld construction\cite{Randall:1999vf} the Friedmann
equation receives an additional term quadratic in the
density\cite{Maartens:2003tw},
\begin{align}
\label{Fried} H^2 = \frac{8\pi}{3 M_P^2} ~ \rho ~ \left(1 +
\frac{\rho}{2\lambda} \right)~, \quad\quad\quad M_P =
\sqrt{\frac{3}{4 \pi}} \frac{M_5^3}{\sqrt{\lambda}}~,
\end{align}
where $M_P$ is the 4D Planck mass and $M_5$ the 5D fundamental
mass and we have set the 4D cosmological constant to zero and
assumed that inflation rapidly makes any dark radiation term
negligible. Eq.~(\ref{Fried}) reduces to the usual Friedmann
equation, $H\propto\sqrt{\rho}$, at sufficiently low energies,
\mbox{$\rho \ll \lambda$}, but at very high energies one has
$H\propto\rho$. Successful BBN requires that the change in the
expansion rate due to the new terms in the Friedmann equation be
sufficiently small at scales $\sim \mathcal{O}$(MeV); this implies
$M_5 \gtrsim 40~\mbox{TeV}$. A more stringent bound, $M_5 \gtrsim
10^5~\mbox{TeV}$, is obtained by requiring the theory to reduce to
Newtonian gravity on scales larger than 1~mm.

We consider the scenario where three heavy right-handed neutrinos
$N_j$, with masses $M_j\,$, are added to the usual particle
content of the minimal supersymmetric standard model. We assume
that the seesaw mechanism is operative in the brane scenario and
gives masses to the light neutrinos, and, for simplicity, we
neglect the dynamics of the heavier sneutrinos
$\widetilde{N}_{2,3}\,$. We also assume that the lightest
right-handed sneutrino $\widetilde{N}_1$ acts as an inflaton with
a potential simply given by the mass term $ V= M_1^2
\widetilde{N}_1^2/2$, i.e., the (chaotic) quadratic potential.

In standard cosmology (SC), the COBE normalisation of the scalar
perturbations requires an inflaton mass $M_1 \sim 10^{13}$~GeV.
This corresponds to an inflaton field value $\sim 3M_P$. These
super-Planckian field values can lead to quantum corrections which
destroy the flatness of the potential necessary for successful
inflation. If we consider the brane modifications to the Friedmann
equation, the COBE normalisation implies\cite{Maartens:1999hf}
that $M_1 \approx 4.5 \times 10^{-5}~M_5$.  This enables inflation
to take place at field values far below $M_{P}$: one estimates
$\widetilde{N}_{1\,i} \approx 3 \times 10^2\,M_5$, which, when
combined with $M_5 \ll 10^{17}$~GeV (required so that inflation
takes place on the high energy regime of BC), implies
$\widetilde{N}_{1\,i} < M_P$. One should also notice that the
inflationary observables, $n_s$, $\alpha_s$ and $r_s$, are well
within the WMAP bounds on these quantities\cite{WMAP}.

\begin{figure}[tb]
\centerline{\epsfxsize=9cm\epsfbox{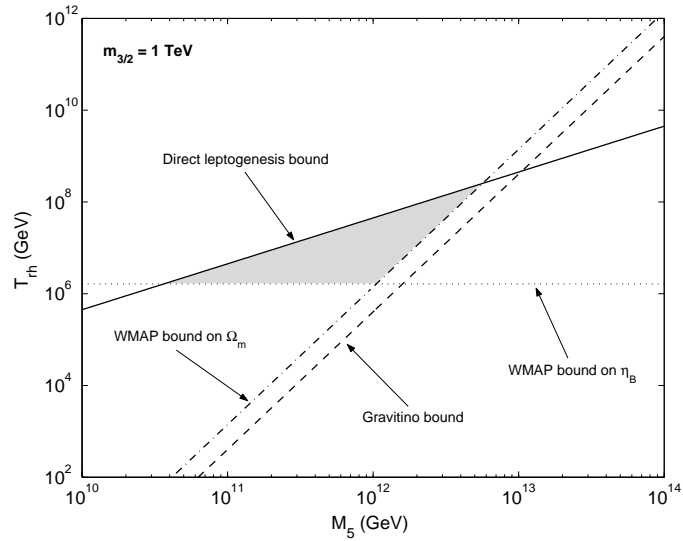}} \caption{The
reheating temperature $T_{rh}$  as a function of $M_5$ for
$m_{3/2}=1~\mbox{TeV}$ as derived from the BBN gravitino
constraints, direct leptogenesis and the WMAP bounds on $\eta_B$ and
$\Omega_m$. \label{fig}}
\end{figure}

At the end of inflation the inflaton field ${\widetilde N}_1$
begins to oscillate coherently around the minimum of the
potential. If $CP$ is not conserved, the decays of ${\widetilde
N}_1$ into leptons, Higgs and the corresponding antiparticles can
produce a net lepton asymmetry. We require $T_{rh} < M_1$, with
$T_{rh}$ being the reheating temperature of the universe, so that
leptogenesis is driven by the decays of the cold sneutrino
inflaton and the produced lepton asymmetry is not washed out by
lepton-number violating interactions mediated by $N_1$ (the case
where leptogenesis is purely thermal, $T_{rh} > M_1$, is not
considered here, but this scenario has been recently investigated
in BC\cite{Bento:2005je}). The lepton-to-photon ratio created is
given by $ \eta_L \sim \epsilon_1 T_{rh}/M_1$, where $\epsilon_1$
denotes the $CP$ asymmetry in the ${\widetilde N}_1$ decays. The
lepton asymmetry produced before the electroweak phase transition
is then partially converted into a baryon asymmetry via the
sphaleron effects\cite{Kuzmin:1985mm}. Taking into account the
observational value for the baryon asymmetry, it is possible to
obtain a lower bound on the reheating temperature, which is
defined by assuming an instantaneous conversion of the inflaton
energy into radiation when the decay width of the inflaton equals
the expansion rate of the universe $H$, $ T_{rh} \gtrsim 1.6
\times 10^6~\mbox{GeV}$.

Another important bound in supersymmetric scenarios comes from
gravitino production. During the reheating gravitinos can be
thermally produced through scatterings in the plasma. If the
gravitinos are unstable and overproduced, their decay products
could put at risk the successful predictions of BBN. If they are
stable particles (which is the case if they are the lightest
supersymmetric particle, LSP) the constraint comes from their
contribution to the dark matter. In SC their abundance is
proportional to $T_{rh}\,$, and constraints from BBN yield a
stringent upper bound\cite{Cyburt:2002uv} on the allowed $T_{rh}$.
In BC, however, their abundance decreases with $T_{rh}\,$, and in
this case we obtain a lower bound on the reheating temperature.

In Fig.~\ref{fig}, we show the allowed region in the $M_5-T_{rh}$
plane, putting together these three constraints, for the case
$m_{3/2}=1$~TeV. In conclusion, for a gravitino mass in the range
$m_{3/2} \simeq 100~{\rm GeV}-1$~TeV, we find that successful BBN
and leptogenesis in this framework require that the 5D Planck mass
is in the range $M_{5} \simeq 10^{10}-10^{13}$~GeV and the
reheating temperature $T_{rh} \simeq 10^{6}-10^{8}$~GeV.


\begin{thebibliography}{0}

\bibitem{WMAP}
D.~N.~Spergel {\it et al.},
Astrophys.\ J.\ Suppl.\  {\bf 148}, 175 (2003);
H.~V.~Peiris {\it et al.},
Astrophys.\ J.\ Suppl.\  {\bf 148}, 213 (2003). %

\bibitem{Bento:2004pz}
For details see  M.~C.~Bento, R.~Gonz\'alez Felipe and
N.~M.~C.~Santos,
  Phys.\ Rev.\ D {\bf 69}, 123513 (2004).
For the study of the mechanism in standard cosmology see:
H.~Murayama, H.~Suzuki, T.~Yanagida and J.~Yokoyama,
Phys.\ Rev.\ Lett.\ \textbf{70}, 1912 (1993);
K.~Hamaguchi, H.~Murayama and T.~Yanagida,
Phys.\ Rev.\ D \textbf{65}, 043512 (2002); 
J.~R.~Ellis, M.~Raidal and T.~Yanagida,
Phys.\ Lett.\ B \textbf{581}, 9 (2004).

\bibitem{Davoudiasl:2004gf}
H.~Davoudiasl, R.~Kitano, G.~D.~Kribs, H.~Murayama and
P.~J.~Steinhardt,
  Phys.\ Rev.\ Lett.\  {\bf 93}, 201301 (2004);
T.~Shiromizu and K.~Koyama,
  JCAP {\bf 0407}, 011 (2004);
  M.~C.~Bento, R.~Gonz\'alez Felipe and N.~M.~C.~Santos,
  Phys.\ Rev.\ D {\bf 71}, 123517 (2005).

\bibitem{Maartens:2003tw}For a review see e.g.
R.~Maartens,
  Living Rev.\ Rel.\  {\bf 7}, 7 (2004).

\bibitem{Randall:1999vf}
L.~Randall and R.~Sundrum,
Phys.\ Rev.\ Lett.\  {\bf 83}, 4690 (1999).

\bibitem{Maartens:1999hf}
R.~Maartens, D.~Wands, B.~A.~Bassett and I.~Heard,
Phys.\ Rev.\ D {\bf 62}, 041301 (2000).

\bibitem{Bento:2005je}
N.~Okada and O.~Seto,
  arXiv:hep-ph/0507279.
M.~C.~Bento, R.~Gonz\'alez Felipe and N.~M.~C.~Santos,
  arXiv:hep-ph/0508213;

\bibitem{Kuzmin:1985mm}
V.~A.~Kuzmin, V.~A.~Rubakov and M.~E.~Shaposhnikov,
Phys.\ Lett.\ B \textbf{155}, 36 (1985).

\bibitem{Cyburt:2002uv}
R.~H.~Cyburt, J.~R.~Ellis, B.~D.~Fields and K.~A.~Olive,
Phys.\ Rev.\ D \textbf{67}, 103521 (2003).

\end{thebibliography}
\end{document}